\documentclass{raa}            

\usepackage{graphicx,times}             
\usepackage{natbib}
\usepackage{amssymb,amsmath}
\bibpunct{(}{)}{;}{a}{}{,}

\begin{document}

   \title{Detection of Gamma-rays from Protostellar Jet in the HH 80-81 System
}

   \volnopage{Vol.0 (20xx) No.0, 000--000}      
   \setcounter{page}{1}          

   \author{Dahai Yan
      \inst{1}
   \and Jianeng Zhou
      \inst{2}
   \and Pengfei Zhang
      \inst{3}
   }

   \institute{Key Laboratory for the Structure and Evolution of Celestial Objects, Yunnan Observatories, Chinese Academy of Sciences, Kunming 650011, China; {\it yandahai@ynao.ac.cn}\\
   \and
      Shanghai Astronomical Observatory, Chinese Academy of Sciences, 80 Nandan Road, Shanghai 200030, China; {\it zjn@shao.ac.cn}\\
        \and
             Department of Astronomy, Key Laboratory of Astroparticle Physics of Yunnan Province, Yunnan University, Kunming 650091, China;  {\it zhangpengfei@ynu.edu.cn}\\
\vs\no
   {\small Received~~20xx month day; accepted~~20xx~~month day}}

\abstract{Considering that the existence of relativistic particles in the protostellar jet has been confirmed by the detection of linearly polarized radio
emission from the HH 80-81 jet, we search for gamma-rays from the HH 80-81 system using ten-year {\it Fermi}-LAT observations.
A significant point-like $\gamma$-ray excess is found in the direction of the HH 80-81 system with Test-Statistic (TS) value $>$100, which is likely produced in the HH 80-81 jet. 
The $\gamma$-ray spectrum extends only to 1 GeV with a photon index of 3.5.
No significant variability is found in the gamma-ray emission.
It is discussed that the properties of HH 80-81 jet suffice for producing the observed $\gamma$-rays.
}

   \authorrunning{Yan et al.}            
   \titlerunning{Gamma-rays from protostellar jet}  

   \maketitle

%
%
\section{Introduction}           
\label{sect:intro}

Radio jets/outflows have been observed in protostars, which are possibly driven by the accretion in the formation of the stars \citep[e.g.,][]{1990ApJ...352..645R,1996RMxAC...4....7R,Marti,1996ApJ...473L.123A,2003ApJ...587..739G,2005ApJ...626..953R}.
In contrast to the relativistic jets in active galactic nuclei (AGNs), the protostellar jets move 
at much smaller velocities which are typically from 100 to 1000 $\rm km\ s^{-1}$ \citep[see][for a review]{2018A&ARv..26....3A}.
The radio emission is usually dominated by the free-free emission from the thermal motions of electrons, which is characterized by a
positive spectral index and no linear polarization \citep[e.g.,][]{2018A&ARv..26....3A}.

Among the protostellar jets, the HH 80-81 jet (located at a distance of 1.7 kpc) is an intriguing object \citep[e.g.,][]{2010Sci}.
The central source in the HH 80-81 system is IRAS 18162-2048, which is identified as a massive B-type protostar \citep[e.g.,][]{2012ApJ...752L..29C}.
The entire radio jet system extends up to $\sim5\ $pc \citep[e.g.,][]{Marti,2010Sci,2017ApJ...851...16R}.
In the central region of the jet, the spectral index of the radio emission is positive, which suggests a dominating thermal emission in the region \citep[e.g.,][]{1996RMxAC...4....7R,2017ApJ...851...16R}.
However, in some knots as well as in the lobes, the radio emissions show negative spectral indices, which suggests an additional non-thermal component
in these regions \citep[e.g.,][]{Marti,2010Sci,2017ApJ...851...16R}.
In particular, the radio emission from the knots located $\sim$0.5 pc from the central source is found to be linearly polarized \citep{2010Sci}.
This clearly confirms the non-thermal origin of the radio emission \citep{2010Sci}.
The non-thermal radio emission is believed to be the synchrotron radiation of relativistic electrons in magnetic field.

Particle acceleration and $\gamma$-ray production in the protostellar jet have been 
studied \citep[e.g.,][]{2007A&A...476.1289A,2010A&A...511A...8B,2013A&A...559A..13M,2016ApJ...818...27R,2017ApJ...851...16R,2019MNRAS.482.4687R}.
The relativistic electrons in the jet would inverse-Compton (IC) scatter optical-ultraviolet (UV) photons to GeV $\gamma$-ray energies.
In a dense material environment, $\gamma$-rays could be produced through relativistic Bremsstrahlung process.
If the protons in the jet are also accelerated accompany with the acceleration of the electrons, the inelastic proton-proton ($pp$) interaction also produce $\gamma$-rays.
However, the detection of $\gamma$-rays from protostellar jet is still lacking.
Motivated by the above arguments, we analyze the {\it Fermi}-LAT data in the direction of IRAS 18162-2048 to search for $\gamma$-rays in the HH 80-81 system.

\section{Data analysis}

In this work, we select 10-year data observed by {\it Fermi}-LAT (from 2008 August 4 to 2018 August 4), covering the energies from 100 MeV to 300 GeV. 
The Pass 8 SOURCE class events within $14^{\circ} \times 14^{\circ}$ region of interest (ROI) 
centered at the position of IRAS 18162-2048 are used \citep{2013arXiv1303.3514A}. 
We use instrument response function (IRF) {\tt P8R3\_SOURCE\_V2}.
The {\it Fermi}-LAT fourth source catalog \citep[4FGL;][]{2019arXiv190210045T} based on the 8-year data is used to construct the background model. 
Galactic and extragalactic diffuse components are modeled by {\tt gll\_iem\_v07.fits} and {\tt iso\_P8R3\_SOURCE\_V2\_v1.txt}, respectively.
We employ the {\tt Fermi-tools} to perform the analysis.
To reduce the contamination from the Earth limb, we exclude the events within zenith angles of $>90^{\circ}$. 
Standard binned likelihood analysis is performed to fit the free parameters in the model,  and the obtained best-fit model is used in next procedure.

\subsection{TS maps and SEDs}

We use {\tt gttsmap} to create $5^{\circ} \times 5^{\circ}$ Test-Statistic (TS) map. 
It is obtained by  moving a putative point source through a grid of locations on the sky and maximizing the likelihood function (-log{\it L}) at each grid point.

\begin{figure*}
 \centering
\includegraphics[width=0.42\columnwidth]{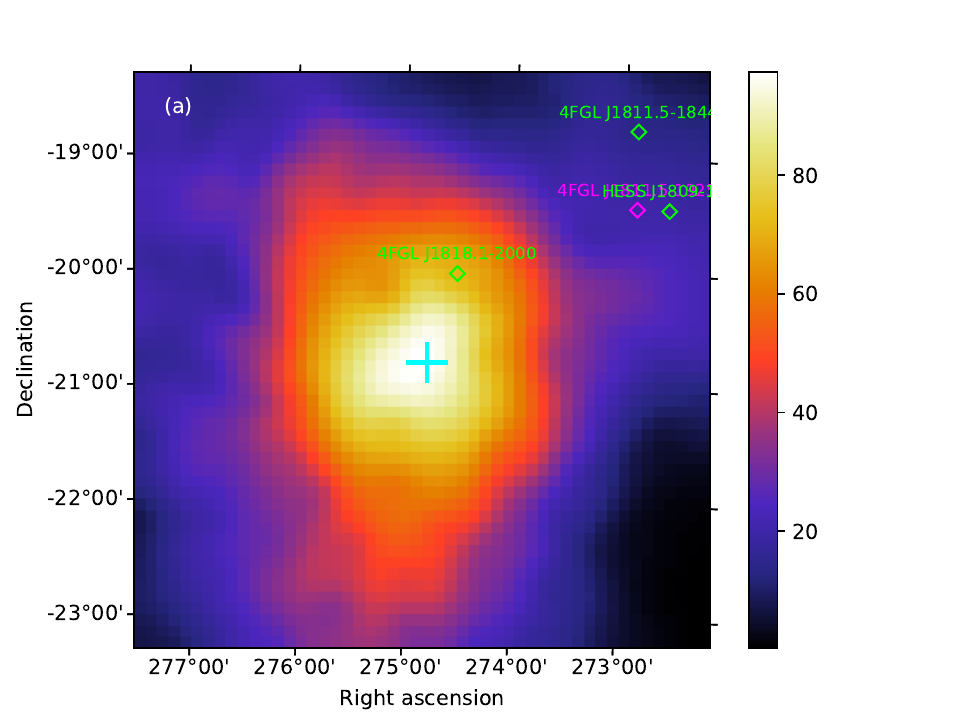}
\includegraphics[width=0.42\columnwidth]{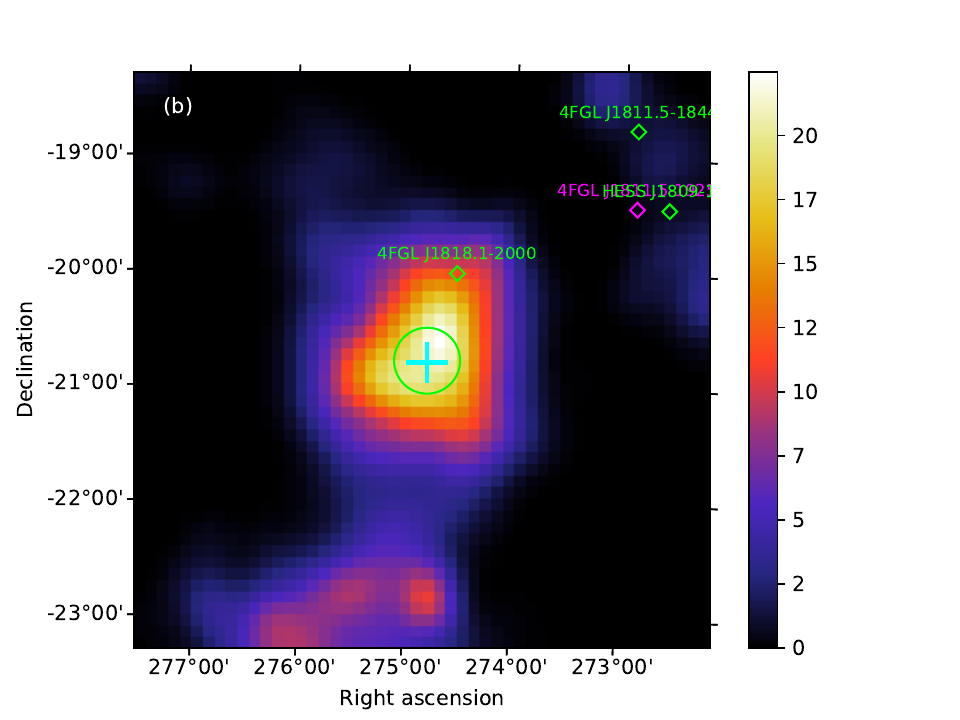}
\includegraphics[width=0.42\columnwidth]{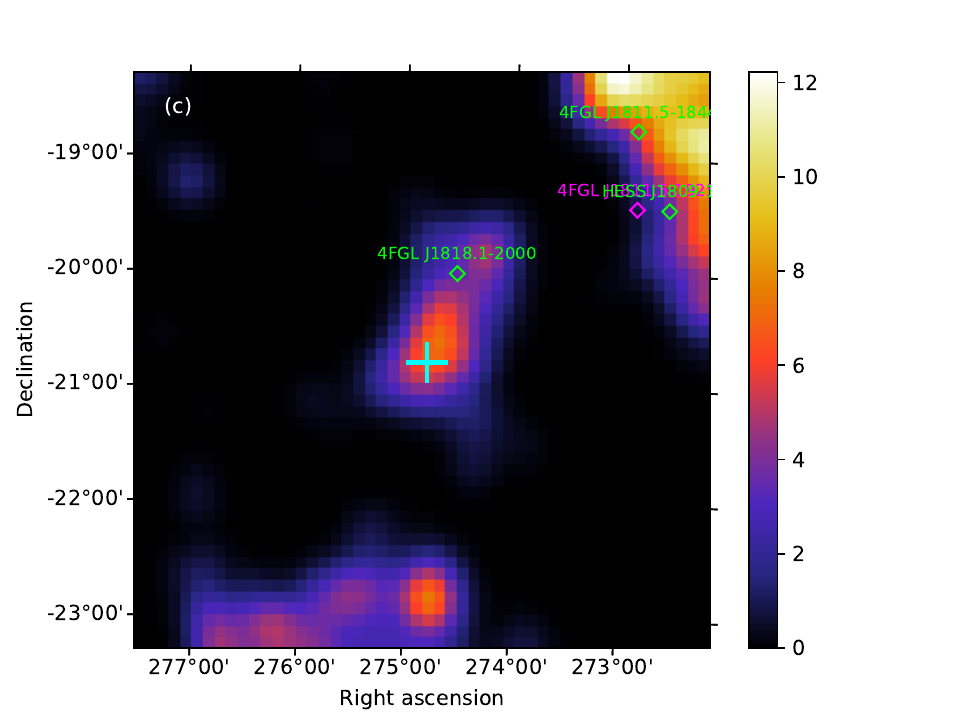}
\includegraphics[width=0.42\columnwidth]{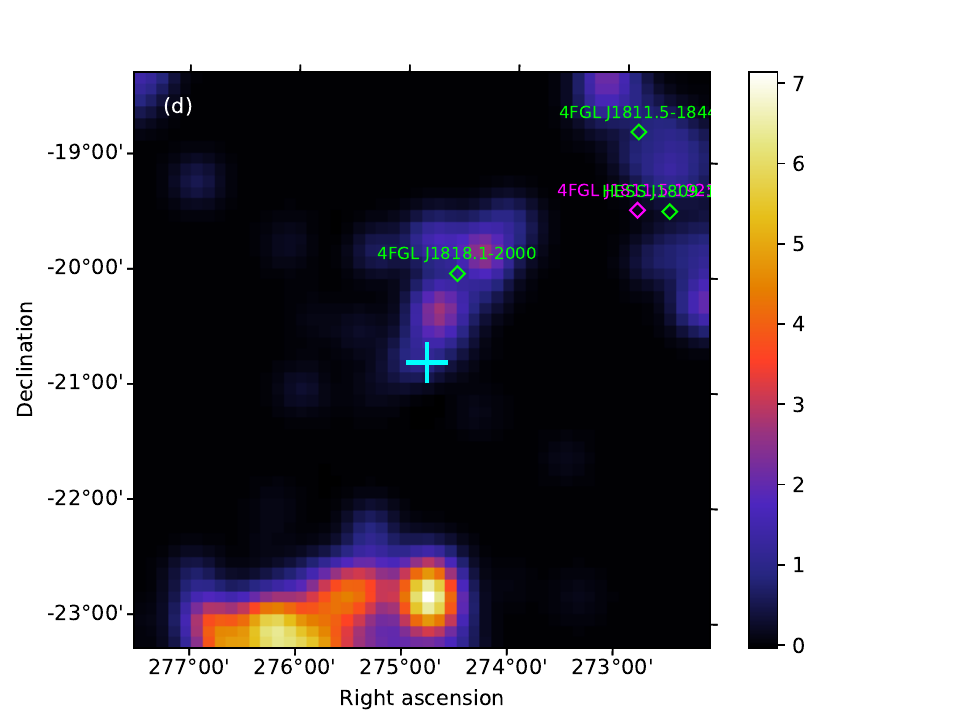}
\caption{$5^{\circ} \times 5^{\circ}$ TS maps of the sky region around IRAS 18162-2048. 
Panels (a)$-$(c) present the maps above 100 MeV, 200 MeV and 300 MeV, respectively.
Panel (d) shows the residual map above 200 MeV considering IRAS 18162-2048 as a power-law point source.
The cyan cross denotes the position of IRAS 18162-2048, and the green circle in (b) is the best-fit position of the $\gamma$-ray excess above 200 MeV. 
All 4FGL sources in this region are labeled as diamonds.}
\label{tsmapSOURCE}
\end{figure*}

In Figure~\ref{tsmapSOURCE}, 
one can see a significant $\gamma$-ray excess in the direction of IRAS 18162-2048.
We add a point-like source at the position of IRAS 18162-2048 to describe this excess. 
The spectrum of the source is assumed to be a power-law form,
\begin{equation}
\frac{dN}{dE}(E)=F_0(\frac{E}{0.1\ \rm GeV})^{-\Gamma_{\gamma}} \ ,
\end{equation}
where the normalization $F_0$ and photon index $\Gamma_{\gamma}$ are free parameters.
Under such assumption, the tool {\tt gtlike} gives TS value of 101 for the emission above 100 MeV (panel a). 
The TS value is reduced to 25 for the emission above 200 MeV (panel b), while above 300 MeV it is only 5 (panel c).
From the residual map (panel d), one can find that the $\gamma$-ray emission can be well described by a point-like source with a power-law spectrum.
The $\gamma$-ray excess has $\Gamma_{\gamma}=3.53 \pm 0.11$ and the flux above 100 MeV $F_{\gamma}=(5.2\pm0.4)\times10^{-8}\ \rm photons\ cm^{-2}\ s^{-1}$. 

\begin{figure}
   \centering
\includegraphics[width=0.68\columnwidth]{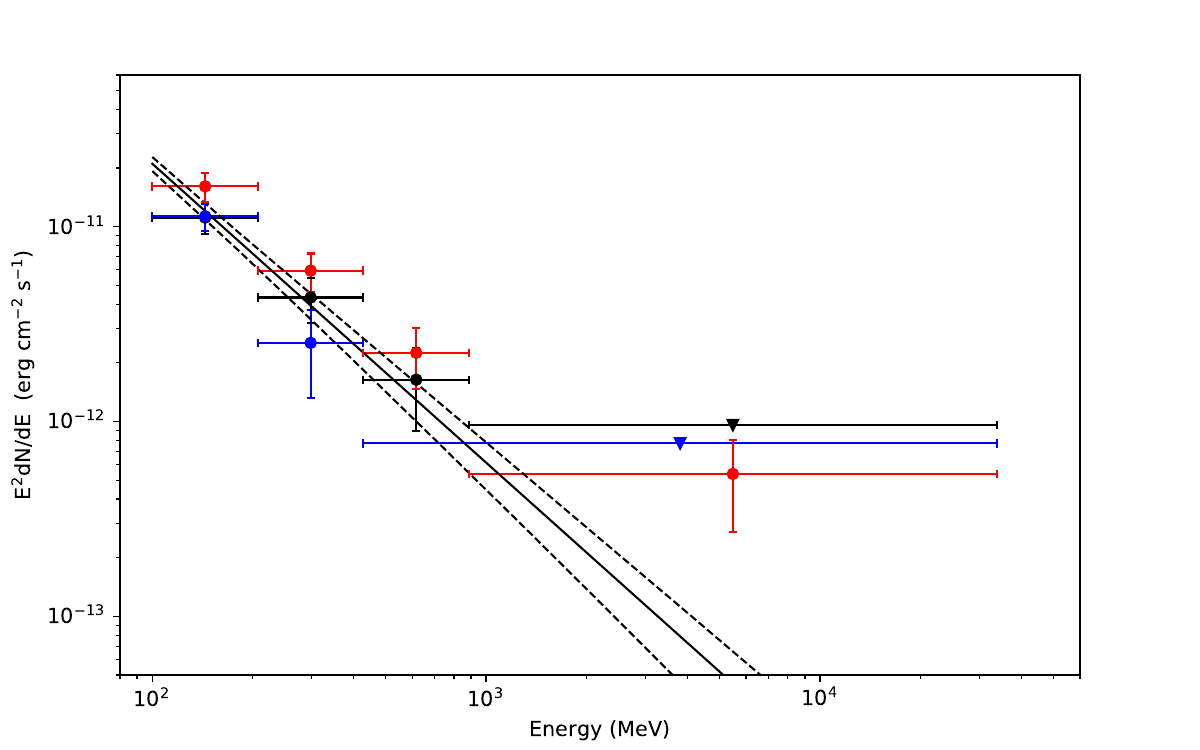}
\caption{
SEDs of the $\gamma$-ray emission from the direction of IRAS 18162-2048. 
The black one is the SED obtained by performing the standard data analysis proceeding;
the red one is that obtained by using the enhanced Galactic diffuse background, and 
the blue one is the SED obtained by using the reduced Galactic diffuse background (see the text for details).
For any bin with TS $<$ 4, upper limit at 95\% confidence level is given (triangle). 
The solid line is the best-fit result to the black points.
Dashed lines represent 1$\sigma$ upper and lower limits for the black points. 
}
\label{sed}
\end{figure}

We then construct spectral energy distribution (SED).
Energies are divided into logarithmic-equivalent bins, and in each bin we perform likelihood analysis to calculate the flux. 
In this step, only the prefactors of all sources are left free (i.e. fix the spectral shapes to those derived before). 
The spectral energy distribution (SED) of the $\gamma$-ray emission is shown in Figure~\ref{sed} (black data).
Again, one can see that the majority of emission is below 1 GeV. 

To investigate the effect of the fluctuation of the diffuse Galactic background (the so-called systematic uncertainty) on the $\gamma$-ray signal,
we change the normalization of the Galactic background component to do the analysis.
The TS value above 100 MeV changes to 76 when the diffuse Galactic background is enhanced by 3\%, and it is 144 when the background is reduced by 3\%.
The derived systematic uncertainty on $F_{\gamma}$  is $\sigma_{\rm syst}=^{+0.45}_{-0.77}\times10^{-8}\ \rm photons\ cm^{-2}\ s^{-1}$. 
The corresponding SEDs are shown in Figure~\ref{sed}  with red and blue circles respectively.
One can find that the effect of the systematic uncertainty on the SED is negligible.

To get accurate position of this $\gamma$-ray emission, we run {\tt gtfindsrc} to optimize the source location. 
This is performed using the photons above 200 MeV, with the consideration of a narrower point spread function (PSF) and lower computation complexity. 
The best-fit position is (R.A.=274.8$^{\circ}$, Decl.=-20.8$^{\circ}$; J2000) with the $1\sigma$ error circle radius of 0.28$^{\circ}$. 
IRAS 18162-2048 is well located in this region.

\subsection{Extension}

In addition to the point-like source model, we also use an uniform disk model as spatial template to investigate the extension of the $\gamma$-ray emission. 
The disk center is set at the position of IRAS 18162-2048, and we vary the disk radius within one degree to calculate the TS values (see Table~\ref{tab:ext}) . 
To clarify a source is extended or not, $\rm TS_{\rm ext} \geq 16$ is usually required \citep{2017ApJ...843..139A} 
where $\rm TS_{\rm ext}$ is defined as $-2({\rm ln}L_{\rm ps}-{\rm ln}L_{\rm ext})$ where $L_{\rm ps}$ and $L_{\rm ext}$ are likelihoods of being a point-like source and an extended source, respectively. 
From Table~\ref{tab:ext}, One can find that the $\gamma$-ray emission does not show extension at an obvious confidence level.

The non-extended feature of the $\gamma$-ray emission disfavors the origins of supernova remnant and molecular cloud.
\footnote{The $\gamma$-ray spectra of molecular clouds below several GeV usually have the photon index of $\sim$2.4 \citep[e.g.,][]{2017A&A...606A..22N}.
The steep $\gamma$-ray spectrum with $\Gamma_{\gamma}=3.53$ cannot be produced by a molecular cloud.}

\begin{table}
\centering
\caption{Significance for testing different radius of uniform disk template. $\Delta L$ is the difference of logarithmic likelihood of disk templates from point source hypothesis.}
\begin{tabular}{c|cccc}
\hline
\hline
Disk radii & 0.1 & 0.3 & 0.5 & 0.7 \\
\hline
-2$\Delta L$ & 0.24 & 0.86 & 0.16 & -1.74 \\
\hline
\end{tabular}
\label{tab:ext}
\end{table}


\subsection{Variability}

\begin{figure}
   \centering
\includegraphics[width=0.68\columnwidth]{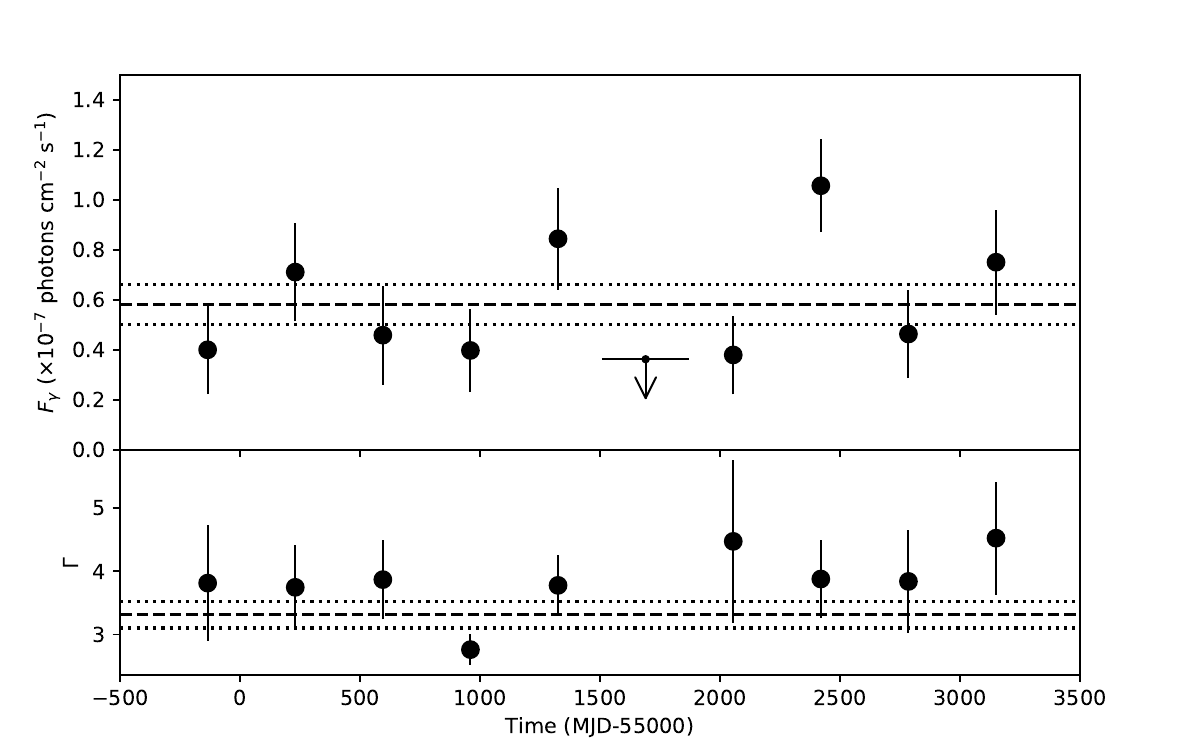}    
\caption{Light curve of the $\gamma$-ray emission from IRAS 18162-2048 direction (top) and the variation of $\Gamma_{\gamma}$ as the time (bottom). The dashed line is the best-fit result to the data with a constant. The region covered by the dotted lines represents the 1 $\sigma$ uncertainty.}
\label{LC}
\end{figure}

We obtain the flux variation during the observation. Light curve is one-year binned (see top panel in Figure~\ref{LC}). 
Since the $\gamma$-ray emission is weak, in each time bin we fix the spectral shapes of background sources  (i.e. only free their prefactors). 
The $\chi^2$ under constant hypothesis is 14.02 with 8 degrees of freedom ({\it d.o.f.}). 
Its corresponding {\it p}-value is 0.08 ($\sim 1.4\sigma$ significance), suggesting no significant variability. 
The variation of $\Gamma_{\gamma}$ is fitted well by a constant of 3.3$\pm$0.2 (see bottom panel in Figure~\ref{LC}).

\section{Cross-checking for Identification}
For the $\gamma$-ray excess region, we exclude known possible $\gamma$-ray sources within the error circle of $0.28^{\circ}$ 
(this circle is obtained from previous {\tt gtfindsrc} performing), by cross-checking known database. 

Totally 315 objects are retrieved in this circle from SIMBAD database\footnote{http://simbad.u-strasbg.fr/simbad/}. 
We carefully check these objects and find that most of them are alternative names of the HH 80-81 system.
Among the rest of the objects (most are stars), no potential $\gamma$-ray emitter is found.
The only $\gamma$-ray candidate 3FGL J1819.5-2045c ($4.5'$ away) has been removed in the {\it Fermi}-LAT fourth source catalog. 
In 4FGL, the nearest known $\gamma$-ray source, 4FGL J1818.1-2000, is about $0.8^{\circ}$ away, which has little impact on the $\gamma$-ray excess (see Figure 1).

The 5th Roma blazar catalog \citep{2009A&A...495..691M,2015Ap&SS.357...75M} contains 3561 sources, 
and the nearest one from IRAS 18162-2048, 5BZQJ1833-2103, has an offset of $3.4^{\circ}$. 
By checking the Australia Telescope National Facility (ATNF) Pulsar Catalog \citep{2005AJ....129.1993M}, no pulsar is found in this circle. 
The Molonglo Reference Catalog (MRC) which is one of the largest homogeneous catalogs of radio sources \citep{1981MNRAS.194..693L,1991Obs...111...72L} do not cover regions within 3 degrees of the Galactic equator. 

\section{Production of the observed Gamma-rays in the HH 80-81 jet} 

Relativistic particles in protostellar jet can produce $\gamma$-rays through a variety of processes, including IC scattering, 
relativistic Bremsstrahlung, and $pp$ interaction \citep[e.g.,][]{2007A&A...476.1289A,2010A&A...511A...8B}.
The present observations cannot determine the radiative mechanism for the $\gamma$-rays.
Nevertheless, the $\gamma$-ray spectrum we obtained can put constraints on the characteristics of the relativistic particles in the jet.

The characteristic cooling time of $pp$ interaction in the hydrogen medium with number density $n_0$ is 
written as \citep[e.g.,][]{2004vhec.book.....A}
\begin{equation}
t_{pp}=(n_0\sigma_{pp}fc)^{-1}\approx5\times10^{7}(n_0/1\ \rm cm^{-3})^{-1}\ yr.
\end{equation}
Here, an average cross-section at high energies of about 40 mb is used, 
and the coefficient of inelasticity $f=0.5$ is adopted (assuming that on average the proton loses about half of its energy per interaction).
For a typical $n_0\sim1000\ \rm cm^{-3}$ in HH 80-81 \citep[e.g.,][]{2018A&ARv..26....3A}, we derive $t_{pp}\sim50000\ $ yr.
The ratio of the mean energy of the produced $\gamma$-ray to the energy of the incident proton is $\sim0.05$ \citep{2006PhRvD..74c4018K}. 
Therefore, the energy of the proton that produces the observed 1 GeV photons through $pp$ interaction is $\sim20$ GeV.
The spectral index of the protons distribution $s_p$ is same as the $\gamma$-ray photon index \citep{2006PhRvD..74c4018K}, i.e., $s_p\approx3.5$.

The cooling time of electrons due to the Bremsstrahlung losses is \citep[e.g.,][]{2004vhec.book.....A}
\begin{equation}
t_{\rm br}\approx4\times10^{7}(n_0/1\ \rm cm^{-3})^{-1}\ yr.
\end{equation}
With $n_0\sim1000\ \rm cm^{-3}$, we get $t_{\rm br}\sim40000\ $ yr.
The energy of the electron that produces the observed 1 GeV photons through Bremsstrahlung is 1 GeV \citep{1970RvMP...42..237B}.
The spectral index of the electron distribution is $s_e\sim\Gamma_{\gamma}\approx3.5$.
In this case, the maximum energy of synchrotron photon is $E_{\rm syn, max} ({\rm eV})\approx2\times10^{-8}(1\ {\rm GeV}/m_ec^2)^2B$.
Considering a magnetic field strength of 0.1 mG \citep{2010Sci}, we have $E_{\rm syn, max}(\rm eV)\sim10^{-5}$. 
This energy range is covered by the current radio observations, and the photon index of the observed radio spectrum is $1.3$ \citep{Marti}. 
From the relativistic electrons distribution, the photon index of the synchrotron emission is expected as $(s_e+1)/2\approx2.3$ \citep{1970RvMP...42..237B}. 
It is noted that this predicted synchrotron spectrum at $\sim$0.01 meV is inconsistent with the radio observations.

The cooling time of relativistic electrons due to the IC scattering is \citep[e.g.,][]{2004vhec.book.....A}
\begin{equation}
t_{\rm IC}=\frac{3m_ec^2}{4\sigma_{\rm T}c\gamma U_{\rm s}}\approx (1/\gamma)(U_{\rm s}/\ \rm erg\ cm^{-3})\ yr,
\end{equation}
where $\gamma$ is the Lorentz factor of the relativistic electrons and $U_{\rm s}$ is the energy density of seed photon field.
The energy of the scattered photons $E_{1}$ is written as $E_1\approx\gamma^2E_{\rm s}$ \citep{1970RvMP...42..237B}, 
where $\gamma$ is the Lorentz factor of the relativistic electrons and $E_{\rm s}$ is the energy of the seed photons.
Protostars are detected as strong infrared sources. Assuming $E_{\rm s}\sim0.01\ $eV (corresponding to the temperature $T\sim100\ $K) \citep{2010A&A...511A...8B},
we can derive $\gamma\approx3\times10^5$ with $E_{1}$=1 GeV.
Considering $U_{\rm s}\sim10^{-12}\ \rm erg\ cm^{-3}$ \citep{2010A&A...511A...8B}, we have $t_{\rm IC}\sim2\times10^6\ $yr for the electrons that produce 1 GeV photons.
The energy of synchrotron photon is calculated by $E_{\rm syn} ({\rm eV})\approx2\times10^{-8}\gamma^2B$, where $B$ is strength of magnetic field.
With $B=0.1\ $mG, the maximum energy of the synchrotron photons is $\approx0.2\ $eV. There is no observed data at this energy range.
In this scenario, $s_e$ is predicted as $s_e=2\Gamma_{\gamma}-1\approx6$.
This very steep spectrum indicates that there is a cut-off at $\gamma_c$ ($\gamma_c<3\times10^5$) in the electron distribution.

As mentioned above, the observed $\gamma$-rays can be produced through IC scattering of relativistic electrons or inelastic $pp$ interaction.
It should be noted that the jet phase timescale in massive protostars is roughly 40000 yr \citep{2012ApJ...753...51G}. 
This lifetime is comparable to $t_{pp}$, but significantly shorter than the characteristic $t_{\rm IC}$ for the electrons that produce 1 GeV photons.

The kinetic luminosity of the jet can be calculated by 
\begin{equation}
L_{\rm j}=\frac{1}{2}\dot{M}_{\rm j}v^2_{\rm j}\ ,
\end{equation}
where $\dot{M}_{\rm j}$ is mass-loss rate in the jet and $v^2_{\rm j}$ is jet velocity.
Taking $\dot{M}_{\rm j}=10^{-6}\ \rm M_{\odot}\ yr^{-1}$ and $v_{\rm j}=1000\ \rm km\ s^{-1}$ \citep[e.g.,][]{2018A&ARv..26....3A}, 
we have $L_{\rm j}\approx3\times10^{35}\ \rm erg\ s^{-1}$. The $\gamma$-ray luminosity ($>$ 100 MeV) is $L_{\gamma}\approx5\times10^{33}\ \rm erg\ s^{-1}$ which is 0.017\% of $L_{\rm j}$.
The energy budget is reasonable.

The detection of the $\gamma$-rays from HH 80-81 not only confirms again that there is a population of relativistic particles in the protostellar jet, 
but also makes the protostellar jet an additional high-energy astrophysics laboratory.
Evidence for strong shocks have been found in the jet of HH 80-81 \citep[e.g.,][]{2019MNRAS.482.4687R}, 
which are produced by the interaction of the jet with the environment. Diffusive Shock Acceleration (DSA) is therefore considered as the mechanism that accelerates particles to high energies \citep[e.g.,][]{2015A&A...582L..13P,2021MNRAS.504.2405A}.
However, details on the acceleration mechanism are still unclear.
The jet properties such as jet ionization fraction and jet temperature which can be probed through detection of molecular emission lines 
are crucial for investigating the acceleration mechanism \citep[e.g.,][]{2021MNRAS.504.2405A}.
Moreover, the detection of non-thermal emissions from radio to X-ray energies is helpful to determine the magnetic field in the jet, 
which is crucial for investigating particle acceleration mechanism and radiative processes.
Follow-up intensive multi-band observations are necessary to reveal the nature of the HH 80-81 jet.

\section{Conclusions}

The clear detection of linearly polarized radio
emission from the HH 80-81 jet \citep{2010Sci} confirms the existence of relativistic particles in the protostellar jet.
$\gamma$-rays from the HH 80-81 jet are therefore expected.
We search for $\gamma$-rays from the HH 80-81 system using 10-year {\it Fermi}-LAT data.
A significant $\gamma$-ray excess is found towards the direction of HH 80-81 system.
After excluding other possible $\gamma$-ray candidates, we suggest that the stable and point-like $\gamma$-ray emission is 
contributed by the HH 80-81 system, which is likely the non-thermal $\gamma$-rays from the HH 80-81 jet.
The spectrum of the $\gamma$-ray emission extends to 1 GeV with a photon index of 3.5.
Moreover, from the aspects of emission mechanism and energy budget, the observed $\gamma$-ray emission can be produced by the HH 80-81 jet.
Our result suggests the protostellar jet as an efficient particle accelerator, which is proposed in theoretical studies \citep[e.g.,][]{2015A&A...582L..13P,2016A&A...590A...8P}.

\section*{Acknowledgements}

We thank the anonymous referee for his/her comments that helped us to improve the quality of our paper. 
We thank Dr. Ruizhi Yang for helpful discussions.
We acknowledge financial supports from the National
Natural Science Foundation of China (NSFC-11803081, NSFC-U1931114, NSFC-U2031205 and NSFC-12163006) 
and the joint foundation of Department of Science and Technology of Yunnan Province and Yunnan University [2018FY001(-003)].
The work of D. H. Yan is also supported by the CAS Youth Innovation Promotion Association and Basic research Program of Yunnan Province (202001AW070013).


%


\begin{thebibliography}{99}

\bibitem[Ackermann et al.(2017)]{2017ApJ...843..139A} Ackermann, M., Ajello, M., Baldini, L., et al.\ 2017, \apj, 843, 139

\bibitem[Aharonian(2004)]{2004vhec.book.....A} Aharonian, F.~A.\ 2004, Very High Energy Cosmic Gamma Radiation: A Crucial Window on the Extreme Universe. Edited by AHARONIAN FELIX A. Published by World Scientific Publishing Co. Pte. Ltd

\bibitem[Atwood et al.(2013)]{2013arXiv1303.3514A} Atwood, W., Albert, A., Baldini, L., et al.\ 2013, arXiv e-prints, arXiv:1303.3514

\bibitem[Anglada et al.(1996)]{1996ApJ...473L.123A}Anglada, G., Rodriguez, L.~F., \& Torrelles, J.~M.\ 1996, \apjl, 473, L123

\bibitem[Anglada et al.(2018)]{2018A&ARv..26....3A} Anglada, G., Rodr{\'\i}guez, L.~F., \& Carrasco-Gonz{\'a}lez, C.\ 2018, \aapr, 26, 3

\bibitem[Araudo et al.(2007)]{2007A&A...476.1289A} Araudo, A.~T., Romero, G.~E., Bosch-Ramon, V., et al.\ 2007, \aap, 476, 1289

\bibitem[\protect\citeauthoryear{Araudo, Padovani, \& Marcowith}{2021}]{2021MNRAS.504.2405A} Araudo A.~T., Padovani M., Marcowith A., 2021, MNRAS, 504, 2405. doi:10.1093/mnras/stab635

\bibitem[Bosch-Ramon et al.(2010)]{2010A&A...511A...8B} Bosch-Ramon, V., Romero, G.~E., Araudo, A.~T., et al.\ 2010, \aap, 511, A8

\bibitem[Blumenthal, \& Gould(1970)]{1970RvMP...42..237B} Blumenthal, G.~R., \& Gould, R.~J.\ 1970, Reviews of Modern Physics, 42, 237

\bibitem[Carrasco-Gonz{\'a}lez et al.(2010)]{2010Sci} Carrasco-Gonz{\'a}lez, C., Rodr{\'\i}guez, L.~F., Anglada, G., et al.\ 2010, Science, 330, 1209

\bibitem[Carrasco-Gonz{\'a}lez et al.(2012)]{2012ApJ...752L..29C} Carrasco-Gonz{\'a}lez, C., Galv{\'a}n-Madrid, R., Anglada, G., et al.\ 2012, \apjl, 752, L29

\bibitem[Garay et al.(2003)]{2003ApJ...587..739G} Garay, G., Brooks, K.~J., Mardones, D., et al.\ 2003, \apj, 587, 739

\bibitem[Guzm{\'a}n et al.(2012)]{2012ApJ...753...51G} Guzm{\'a}n, A.~E., Garay, G., Brooks, K.~J., et al.\ 2012, \apj, 753, 51

\bibitem[Kelner et al.(2006)]{2006PhRvD..74c4018K} Kelner, S.~R., Aharonian, F.~A., \& Bugayov, V.~V.\ 2006, \prd, 74, 034018

\bibitem[The Fermi-LAT collaboration(2019)]{2019arXiv190210045T} The Fermi-LAT collaboration\ 2019, arXiv e-prints, arXiv:1902.10045

\bibitem[Large et al.(1981)]{1981MNRAS.194..693L} Large, M.~I., Mills, B.~Y., Little, A.~G., et al.\ 1981, \mnras, 194, 693

\bibitem[Large et al.(1991)]{1991Obs...111...72L} Large, M.~I., Cram, L.~E., \& Burgess, A.~M.\ 1991, The Observatory, 111, 72

\bibitem[Laing et al.(1983)]{1983MNRAS.204..151L} Laing, R.~A., Riley, J.~M., \& Longair, M.~S.\ 1983, \mnras, 204, 151

\bibitem[Marti et al.(1993)]{Marti}Mart{\'i}, J., Rodr{\'i}guez, L. F., \& Reipurth, B. 1993, ApJ, 416, 208

\bibitem[Manchester et al.(2005)]{2005AJ....129.1993M} Manchester, R.~N., Hobbs, G.~B., Teoh, A., et al.\ 2005, \aj, 129, 1993

\bibitem[Munar-Adrover et al.(2013)]{2013A&A...559A..13M} Munar-Adrover, P., Bosch-Ramon, V., Paredes, J.~M., et al.\ 2013, \aap, 559, A13

\bibitem[Massaro et al.(2015)]{2015Ap&SS.357...75M} Massaro, E., Maselli, A., Leto, C., et al.\ 2015, \apss, 357, 75

\bibitem[Massaro et al.(2009)]{2009A&A...495..691M} Massaro, E., Giommi, P., Leto, C., et al.\ 2009, \aap, 495, 691

\bibitem[Neronov et al.(2017)]{2017A&A...606A..22N} Neronov, A., Malyshev, D., \& Semikoz, D.~V.\ 2017, \aap, 606, A22

\bibitem[Padovani et al.(2015)]{2015A&A...582L..13P} Padovani, M., Hennebelle, P., Marcowith, A., et al.\ 2015, \aap, 582, L13

\bibitem[Padovani et al.(2016)]{2016A&A...590A...8P} Padovani, M., Marcowith, A., Hennebelle, P., et al.\ 2016, \aap, 590, A8

\bibitem[Rodr{\'\i}guez et al.(1990)]{1990ApJ...352..645R} Rodr{\'\i}guez, L.~F., Ho, P.~T.~P., Torrelles, J.~M., et al.\ 1990, \apj, 352, 645

\bibitem[Rodriguez(1996)]{1996RMxAC...4....7R} Rodriguez, L.~F.\ 1996, Revista Mexicana De Astronomia Y Astrofisica Conference Series, 7

\bibitem[Rodr{\'\i}guez et al.(2005)]{2005ApJ...626..953R} Rodr{\'\i}guez, L.~F., Garay, G., Brooks, K.~J., et al.\ 2005, \apj, 626, 953

\bibitem[Rodr{\'\i}guez-Kamenetzky et al.(2016)]{2016ApJ...818...27R} Rodr{\'\i}guez-Kamenetzky, A., Carrasco-Gonz{\'a}lez, C., Araudo, A., et al.\ 2016, \apj, 818, 27

\bibitem[Rodr{\'\i}guez-Kamenetzky et al.(2017)]{2017ApJ...851...16R} Rodr{\'\i}guez-Kamenetzky, A., Carrasco-Gonz{\'a}lez, C., Araudo, A., et al.\ 2017, \apj, 851, 16

\bibitem[Rodr{\'\i}guez-Kamenetzky et al.(2019)]{2019MNRAS.482.4687R} Rodr{\'\i}guez-Kamenetzky, A., Carrasco-Gonz{\'a}lez, C., Gonz{\'a}lez-Mart{\'\i}n, O., et al.\ 2019, \mnras, 482, 4687

\end{thebibliography}

\label{lastpage}

\end{document}